\documentclass[article]{aa} 
\pdfoutput=1

\usepackage{ragged2e}
\usepackage{graphicx}
\usepackage{txfonts}
\usepackage[]{natbib}
\usepackage{multirow}
\usepackage{amsmath}
\usepackage[colorlinks=true,linkcolor=blue,citecolor=blue]{hyperref}
\def\deg{$^{\circ}$}

\newcommand{\elodie}[1]{#1}

\begin{document} 

\title{First resolved observations of a highly asymmetric debris disc around HD\,160305 with VLT/SPHERE \thanks{Based  on  observations  made  with  ESO  Telescopes  at  the  Paranal Observatory under programs ID 95.C-0298 and 97.C-0865.}
\thanks{The reduced images as FITS files are only available in electronic form at the CDS via anonymous ftp to cdsarc.u-strasbg.fr (130.79.128.5) or via http://cdsweb.u-strasbg.fr/cgi-bin/qcat?J/A+A/}
}
\author{Clément~Perrot\inst{\ref{lesia},\ref{ifa},\ref{npf}} \and 
Philippe~Thebault\inst{\ref{lesia}}\and 
Anne-Marie~Lagrange\inst{\ref{ipag}}\and
Anthony~Boccaletti\inst{\ref{lesia}}\and 
Arthur~Vigan\inst{\ref{lam}}\and 
Silvano~Desidera\inst{\ref{inaf}}\and
Jean-Charles~Augereau\inst{\ref{ipag}}\and 
Mickael~Bonnefoy\inst{\ref{ipag}}\and 
\'Elodie~Choquet\inst{\ref{lam}}\and 
Quentin~Kral\inst{\ref{lesia}}\and 
Alan~Loh\inst{\ref{lesia}}\and
Anne-Lise~Maire\inst{\ref{belge},\ref{mpia}}\and 
François~Ménard\inst{\ref{ipag}}\and 
Sergio~Messina\inst{\ref{inaf2}}\and 
Johan~Olofsson\inst{\ref{mpia},\ref{ifa},\ref{npf}} \and 
Raffaele~Gratton\inst{\ref{inaf}}\and
Beth~Biller\inst{\ref{edin},\ref{mpia}}\and 
Wolfgang~Brandner\inst{\ref{mpia}}\and
Esther~Buenzli\inst{\ref{mpia},\ref{ethz}}\and
Gaël~Chauvin\inst{\ref{ipag},\ref{chili}}\and
Anthony~Cheetham\inst{\ref{geneve}}\and
Sebastien~Daemgen\inst{\ref{ethz}}\and
Philippe~Delorme\inst{\ref{ipag}}\and
Markus~Feldt\inst{\ref{ipag}}\and
Eric~Lagadec\inst{\ref{oca}}\and
Maud~Langlois\inst{\ref{cral},\ref{lam}}\and
Justine~Lannier\inst{\ref{ipag}}\and
Dino~Mesa\inst{\ref{dino},\ref{inaf}}\and
David~Mouillet\inst{\ref{ipag}}\and
Sébastien~Peretti\inst{\ref{geneve}}\and
Pierre~Janin-Potiron\inst{\ref{ipag}}\and
Graeme~Salter\inst{\ref{lam}}\and
Elena~Sissa\inst{\ref{inaf}}\and
Alain~Roux\inst{\ref{ipag}}\and
Marc~Llored\inst{\ref{lam}}\and
Jean-Tristan~Buey\inst{\ref{lesia}}\and
Alexei~Pavlov\inst{\ref{mpia}}\and
Luc~Weber\inst{\ref{geneve}}\and
Cyril~Petit\inst{\ref{onera}} }
\institute{LESIA, Observatoire de Paris, Université PSL, CNRS, Sorbonne Université, Univ. Paris Diderot, Sorbonne Paris Cité, 5 place Jules Janssen, 92195 Meudon, France\label{lesia}
\and Instituto de Física y Astronomía, Facultad de Ciencias, Universidad de Valparaíso, Av. Gran Bretaña 1111, Valparaíso, Chile\label{ifa}
\and N\'ucleo Milenio Formaci\'on Planetaria - NPF, Universidad de Valpara\'iso, Av. Gran Breta\~na 1111, Valpara\'iso, Chile\label{npf}
\and Univ. Grenoble Alpes, CNRS, IPAG, F-38000 Grenoble, France\label{ipag}
\and Aix Marseille Univ, CNRS, CNES,  LAM, Marseille, France\label{lam}
\and INAF - Osservatorio Astronomico di Padova, Vicolo dell Osservatorio 5, 35122, Padova, Italy\label{inaf}
\and INAF - Osservatorio Astrofisico di Catania, Via S.Sofia 78, 95123, Catania, Italy\label{inaf2}
\and Max Planck Institute for Astronomy, K\"onigstuhl 17, D-69117 Heidelberg, Germany\label{mpia}
\and Institute for Astronomy, University of Edinburgh, EH9 3HJ, Edinburgh, UK\label{edin}
\and Institute for Particle Physics and Astrophysics, ETH Zurich, Wolfgang-Pauli-Strasse 27, 8093 Zurich, Switzerland\label{ethz} 
\and Unidad Mixta Internacional Franco-Chilena de Astronom\'{i}a, CNRS/INSU UMI 3386 and Departamento de Astronom\'{i}a, Universidad de Chile, Casilla 36-D, Santiago, Chile\label{chili}
\and Geneva Observatory, University of Geneva, Chemin des Mailettes 51, 1290 Versoix, Switzerland\label{geneve}
\and Universite Cote d’Azur, OCA, CNRS, Lagrange, France\label{oca}
\and CRAL, CNRS, Universite Lyon 1, ENS, 9 avenue Charles Andre, 69561 Saint Genis Laval, France\label{cral}
\and STAR Institute, University of Liège, Allée du Six Août 19c, B-4000 Liège, Belgium\label{belge}
\and INCT, Universidad De Atacama, calle Copayapu 485, Copiapò, Atacama, Chile\label{dino}
\and DOTA, ONERA, Université Paris Saclay, F-91123, Palaiseau France\label{onera}
} 
\date{Received 19 November 2018 / Accepted 15 April 2019}
\offprints{Clément Perrot, \email{clement.perrot 'at' obspm.fr} or \email{clement.perrot 'at' uv.cl} }

\titlerunning{First resolved observations of a highly asymmetric debris disc around HD\,160305 with VLT/SPHERE}

\authorrunning{Perrot et al.}

\abstract
        {Direct imaging of debris discs gives important information about their nature, their global morphology, and allows us to identify specific structures possibly in connection with the presence of gravitational perturbers. It is the most straightforward technique to observe planetary systems as a whole.
    }
        {We present the first resolved images of \elodie{the} debris disc around the young F-type star HD\,160305, \elodie{detected in scattered light} using the VLT/SPHERE instrument in the near infrared.
    }
        {We used \elodie{a} post-processing method based on angular differential imaging \elodie{and} synthetic images of debris discs produced with a \elodie{disc modelling code}  (\textit{GRaTer}) to constrain the main characteristics of the disc around HD\,160305. All \elodie{of} the point sources in the field of the IRDIS camera \elodie{were} analysed with an astrometric tool to determine whether they are bound objects or background stars.
    }
        {We detect a very inclined ($\sim$ 82$^{\circ}$) ring-like debris disc located at a stellocentric distance of about 86\,au (deprojected width $\sim$27\,au). The disc displays a brightness asymmetry between the two sides of the major axis, as can be expected from scattering properties of dust grains. We derive an anisotropic scattering factor g$>$0.5. A second right-left asymmetry is also observed with respect to the minor axis. We measure a surface brightness ratio of 0.73\,$\pm$\,0.18 between the bright and the faint sides. Because of the low signal-to-noise ratio (S/N) of the images we cannot easily discriminate between several possible explanations for this left-right asymmetry, such as perturbations \elodie{by} an unseen planet, the aftermath of the breakup of a massive planetesimal, or \elodie{the} pericenter glow effect due to an eccentric ring. Two epochs of observations allow us to reject the companionship hypothesis for the 15 point sources present in the field.
    }
    {}
 
\maketitle

\section{Introduction}
\label{intro}

In the current model of planetary system formation, a circumstellar disc evolves from the protoplanetary disc phase \citep[optically thick, dominated by gas;][]{2011ARA&A..49...67W} to the debris disc phase \citep[optically thin, with low or no gas content;][]{2018ARA&A..56..541H}. 
This inside-out process lasts typically a few million years during which the primordial gas dissipates, leaving only planets, if already formed, and planetesimal belts \citep{2007ApJ...662.1067H}. 
The latter trigger the production of a new generation of collisionally induced dust grains, potentially detectable in scattered light and thermal emission \citep[see review of][]{2018ARA&A..56..541H}.\\

With the advent of dedicated instruments for high-contrast imaging (first the Hubble Space Telescope, see \textit{e.g.} \citealt{1999ApJ...513L.127S, 2007ApJ...661L..85K,2016ApJ...817L...2C} and more recently the Spectro-Polarimetric High-contrast Exoplanet REsearch (SPHERE), \citealt{2019arXiv190204080B} and the Gemini Planet Imager (GPI), \citealt{2014PNAS..11112661M}), a variety of morphologies were discovered among debris discs.
Several imaged discs present pronounced azimuthal asymmetries (HD\,106906, \citealt{2015ApJ...814...32K,2016A&A...586L...8L}, HD\,61005, \citealt{2007ApJ...671L.165H,2016A&A...591A.108O} or GSC\,07396-00759, \citealt{2018A&A...613L...6S}). 
Other discs display several concentric belts, like HD\,131835 \citep{2017A&A...601A...7F} or the hybrid disc \citep[or potentially shielded disc of secondary origin,][]{2018arXiv181108439K} HD\,141569 \citep{2016A&A...590L...7P}. In some cases, very complex morphologies have been identified, like the very fast moving structures in the debris disc of AU Mic \citep{2015Natur.526..230B,2018A&A...614A..52B}. For most cases, the interaction between the disc and a potential massive companion has been invoked to explain these spatial structures \citep[e.g.][]{2016ApJ...827..125L}. 
However, apart from the particular case of $\beta$ Pictoris in which the planet b is known to have induced a disc feature \citep[a warp,][]{1997MNRAS.292..896M,2001A&A...370..447A,2010Sci...329...57L}, the simultaneous detection of planet(s) and a disc is very uncommon in scattered light. \citet{2016ApJ...827..125L} showed that \elodie{a planet of a} few Earth masses \elodie{only is} enough to create such asymmetries, \elodie{yet such low-mass objects} are beyond the reach of the current high-contrast imaging instruments.
An additional example is the detected moving companion of Fomalhaut \citep{2005Natur.435.1067K,2008Sci...322.1345K}, however, its \elodie{planetary} nature is still discussed \citep{2015MNRAS.448..376N,2017MNRAS.468.4018P}.
For most imaged asymmetric discs, where no companion has been detected (yet), other explanations than the planet-sculpting scenario might also be possible. Such scenarios include the breakup of a massive planetesimal \citep{2015A&A...573A..39K}, or the interaction between dust and gas due for example to\textit{} the photoelectric instability of the dust \citep{2013Natur.499..184L,2018MNRAS.477.5191R}.\\

HD\,160305 (HIP\,86598) is a F9V type star located at 65.51\,$\pm$\,0.23\,pc \citep[previous value from Hipparcos: 72.46\,$\pm$\,4.5\,pc,][]{2007A&A...474..653V}, with a proper motion of -2.02\,$\pm$\,0.07\,mas.yr$^{-1}$ in RA and -65.86\,$\pm$\,0.06\,mas.yr$^{-1}$ in Dec \citep{2018A&A...616A...1G}.
The star was first proposed as a member of the $\beta$ Pictoris moving group (BPMG) by \citet{2011MNRAS.411..117K}. This membership was \elodie{later} confirmed by \citet{2013ApJ...762...88M}, \elodie{but also contested} by \citet{2012AJ....144....8S}, who suggested that HD\,160305 is a member of the Sco-Cen complex. However, we note that the star is quite far from this region in the sky, about 10\deg\,apart from the closest group boundary as defined in \citet{2016MNRAS.461..794P}.
The most recent analyses by \citet{2017A&A...600A..83M} and \citet{2018MNRAS.475.2955L} further support the BPMG membership.
Few other mild outliers are present within the bona-fide members, especially for stars with debris discs. 
Indirect age indicators like lithium and X-ray emission, although of limited sensitivity for an F9V star, are compatible with this membership as well as the isochrone fitting \citep{2015MNRAS.454..593B}. Finally, using the recent kinematic data from Gaia Data Release 2 (DR2) \citep{2018A&A...616A...1G} and the BANYAN $\Sigma$ tool\footnote{Bayesian Analysis for Nearby Young AssociatioNs: http://www.exoplanetes.umontreal.ca/banyan/banyansigma.php.} from \citet{2018ApJ...856...23G}, we obtain a probability of 99.4\% that the star is a member of the BPMG.
We \elodie{thus} adopt the age of the BPMG in the following: 23\,$\pm$\,3\,Myrs \citep{2014MNRAS.445.2169M}.
\citet[PhD Thesis,][]{schneiderphdthesis} reports a projected rotational velocity of the star vsini\,=\,37\,km.s$^{-1}$. Combined with the stellar rotation period P\,=\,1.336\,$\pm$\,0.008 days \citep{2017A&A...600A..83M}, we can derive a stellar rotation axis of $i_\star$\,=\,58\deg$^{+18}_{-10}$ (Appendix \ref{sini}). \citet{2012AcA....62...67K} also reported a small stellar variability at visible wavelengths, $\Delta$I\,=\,0.041 mag and $\Delta$V\,=\,0.047 mag. 
A search in Gaia DR2 did not reveal any companions within 20 arcmin \citep[that is $\sim$0.3 pc at the distance of the star,][]{2018A&A...616A...1G}. Gaia would reveal any companion with a G-band magnitude < 19.5, which implies a mass larger than $\sim$ 0.15 M$_\odot$ at the distance of HD\,160305. While not fully complete, there is little room for stellar companions.
Moreover, no gas was detected in this system, as reported by \citet{2015ApJ...814...42M}.\\

The WISE\footnote{Wide-Field Infrared Survey Explorer.} survey \citep{2013ApJS..208....9P} identified an infrared excess at 2.9$\sigma$ based on the Ks\,-\,W4 colour, confirmed by \citet{2016ApJ...826..123M} with Herschel observations at 70\,$\mu$m and 160\,$\mu$m. Using a single temperature spectral energy distribution (SED) fitting of the far-infrared excess, \citet{2016ApJ...826..123M} derived a first estimation of the disc's properties, with a radius at 58\,$\pm$\,13\,au, a disc temperature of 43\,$\pm$\,5\,K, and a fraction of dust luminosity with respect to the bolometric luminosity of the star of $1.2\,\pm\,0.3\times10^{-4}$. 
This system \elodie{had} never been \elodie{imaged} to date, neither in the submillimetric in thermal emission nor in scattered light.
The \elodie{SED-inferred} disc radius corresponds to an angular size between 0.6\arcsec\,and 1\arcsec, which is between the inner working angle and the field of view of SPHERE.
Spatially resolving discs provides direct measurements of \elodie{their} radius and dust extent (\textit{e.g.} ring versus extended disc) and allows us to break the degeneracies between the temperature and the distance to the star of the dust \elodie{in SED fits}.\\

In this paper we present the first resolved images of the faint disc around HD\,160305, obtained in the near infrared with the SPHERE instrument. \elodie{The} observations are described in Sect. \ref{obs}. The companionship analysis of the various point sources detected in our image is presented in Sect. \ref{cc}. Section \ref{disk}  presents the modelling of the disc. We then discuss in Sect. \ref{disk} the different scenarios that could explain the origin of the observed disc asymmetry.\\

\section{Observations}
\label{obs}

\begin{figure*}[ht] 
\includegraphics[width=1.0\textwidth,clip]{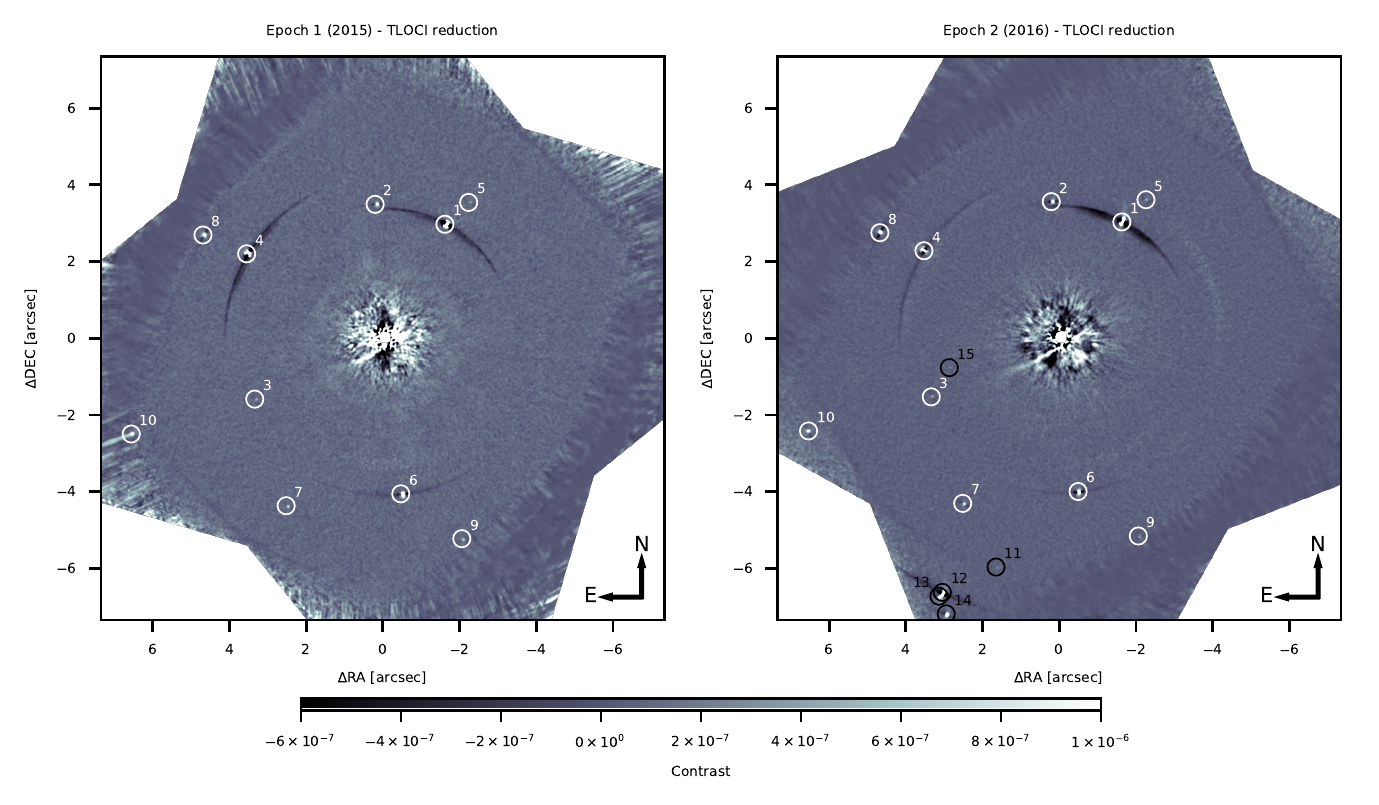}
\normalsize
\caption{IRDIS images of the first and second epochs, reduced with the TLOCI algorithm. North is up and east is left. Point sources are indicated with circles and numeric labels. Whites circles are used for point sources detected at both epochs and black circles for point sources detected at the second epoch only.}
\label{irdis2015-2016}
\end{figure*}

HD\,160305 was observed twice with SPHERE \citep{2019arXiv190204080B} as part of the guaranteed time observation (GTO) programme: SpHere INfrared survey for Exoplanets \citep[SHINE;][]{2017sf2a.conf..331C}.\\

On the first epoch (May 13$^{}$, 2015) we used the standard IRDIFS mode combining simultaneously the Integral Field Spectrometer \citep[IFS,][]{2008SPIE.7014E..3EC} in the YJ mode (0.95~-~1.35\,$\mu$m) and the Infra-Red Dual-beam Imager and Spectrograph \citep[IRDIS,][]{2008SPIE.7014E..3LD} \elodie{with the} dual-band filters H2-H3 \citep[$\lambda_{H2}$\,=\,1.593\,$\mu$m, $\lambda_{H3}$\,=\,1.667\,$\mu$m, $\delta\lambda$\,=\,53\,nm,][]{2010SPIE.7735E..2XV}.
Sky conditions were good but we experienced some open loop \elodie{interruption}s of the adaptive optics \elodie{control} due to instabilities that required frame sorting during the data analysis. The disc was barely detected in this first data set as a very faint structure. 
To \elodie{improve} the detection of the disc, we used a hybrid configuration for the second epoch (May 23$^{}$, 2016) where IRDIS was operated in classical imaging mode with the broad-band filter BB\_H \citep[$\lambda_c$\,=\,1.625\,$\mu$m, $\delta\lambda$\,=\,290\,nm,][]{2014SPIE.9147E..9PL} in both spectral channels, and IFS still set up in the YJ mode. 
Both observations were done with the 92.5\,mas radius Apodized Lyot Coronagraph \citep[APLC;][]{2011ExA....30...59G}. The fields of view of IRDIS and IFS are 11\arcsec\,$\times$\,12.5\arcsec\,and 1.73\arcsec\,$\times$\,1.73\arcsec\,respectively, with a pixel-scale of 12.242\,$\pm$\,0.062\,mas (first epoch) and 12.247\,$\pm$\,0.017\,mas (second epoch) for IRDIS, and 7.46$\,\pm$\,0.02\,mas for IFS \citep{2016SPIE.9908E..34M}. Using \elodie{the} NGC6380 and NGC3603 clusters as astrometric references, we measured \elodie{the} true north orientation of 1.712\,$\pm$\,0.063\deg\,and 1.675\,$\pm$\,0.080\deg\,for, respectively, the first and \elodie{the} second epoch. The setup and parameters of observations are summarized in Table \ref{logobs}.\\

All SHINE observations follow a similar sequence: a first series of images of the star taken out of the coronagraphic mask and with a neutral density to avoid saturation. These images are used to calibrate the flux in the coronagraphic images (flux). 
Then, several coronagraphic data were acquired in waffle mode, which consists in applying a periodic modulation to the deformable mirror to create four crosswise spots around the star \citep{2013aoel.confE..63L}. These spots allow us to determine precisely the position of the star behind the coronagraph (centring). 
Afterwards, a long series of deep coronagraphic exposures (science) \elodie{was} acquired, followed by a second series of coronagraphic images in waffle mode, a second \elodie{unocculted} star image to assess the flux variability during the sequence, and finally a few sky frames to subtract the bias, the dark, and the background from the deep coronagraphic exposures. All the remaining calibrations (darks, flats, wavelength calibration for IFS) are carried out in daytime. A GTO run comes with astrometric calibrator observations to determine the pixel scale and the true north orientation.\\

Both IRDIS and IFS data were reduced using the SPHERE data reduction and handling (DRH) pipeline \citep{2008ASPC..394..581P}, hosted at the SPHERE data centre (DC) in Grenoble\footnote{http://sphere.osug.fr/spip.php?rublique16\&lang=en.} \citep{2017sf2a.conf..347D}. 
The reduction applied to the data follows the standard procedure of SHINE targets including background subtraction, bad pixel and flat-field corrections, and centring of the coronagraphic frames using the "waffle" mode observation. \elodie{The IRDIS acquisitions included a dithering pattern to help with bad pixel rejection.} 
Additional custom routines were also used (frame centring and sorting), in particular for the IFS data reduction \citep{2015A&A...576A.121M}. 
\elodie{The} IFS data \elodie{were} spectrally calibrated owing to internal calibration laser lines. The true-north (TN), the pixel-scale, and the distortion correction \elodie{were} determined for both epochs using the same method described by \citet{2016SPIE.9908E..34M}. 
Sorting of coronagraphic frames was required for both epochs to remove open adaptive optics loops and poor-quality frames (poor adaptive optics corrections). A total of ten out of 64 frames, and 25 out of 144 frames, were rejected for the first and second epochs, respectively. 
Each data set is composed of a temporal and spectral cube of coronagraphic images, and of the unsaturated images of the star (point-spread function, PSF).\\

After the standard reduction, the data are post-processed with the \textit{SpeCal} software \citep{2018A&A...615A..92G}, hosted in the DC, which allows us to use a variety of post-processing methods based on angular differential imaging \citep[ADI,][]{2006ApJ...641..556M}, spectral differential imaging \citep[SDI,][]{1987PASP...99.1344S}, or a combination of both. 
The reductions revealed the presence of several point sources and a resolved, though very faint, disc around the star.
In this paper, we present the results \elodie{obtained with} post-processing based on the ADI method. We used the Template Locally Optimized Combination of Images \citep[TLOCI,][]{2014SPIE.9148E..0UM} and the Karhunen-Loève Image Projection \citep[KLIP,][]{2012ApJ...755L..28S} algorithms, which provide, in this case, the best reductions for the point sources analysis and \elodie{for} the disc analysis, respectively (Figs. \ref{irdis2015-2016} and \ref{disksnr}). For both epochs, the KLIP reductions are performed in a 2.5\arcsec\,radius field of view and a truncation of ten modes. The TLOCI reductions are performed for the entire images, using the same parameters as those used for the standard reduction of SHINE \elodie{data}, as defined in \citet{2018A&A...615A..92G}:
Perc = 0.1, the minimum throughput allowed for a point source;
$\Delta r(red)$ = 1.5$\lambda/D$, the width of the area where speckles are suppressed;
$\Delta r(opt)$ = 1.5$\lambda/D$, the width of the area where the TLOCI coefficients are calculated;
$N_{A}$ = 20, \elodie{the} size in PSF width-units of the area where speckles are suppressed. \\  

\elodie{We} note that previous observations of the star were made with VLT/NaCo on the 1 and the 25$^{}$ of June 2009 in the Ks band. Those observations were performed without coronagraph, saturating the central star to enhance the dynamical range of the data. Yet, the data quality was not sufficient to detect the disc.

\section{Point sources analysis}
\label{cc}

\begin{figure}[ht] 
\includegraphics[width=0.49\textwidth,clip]{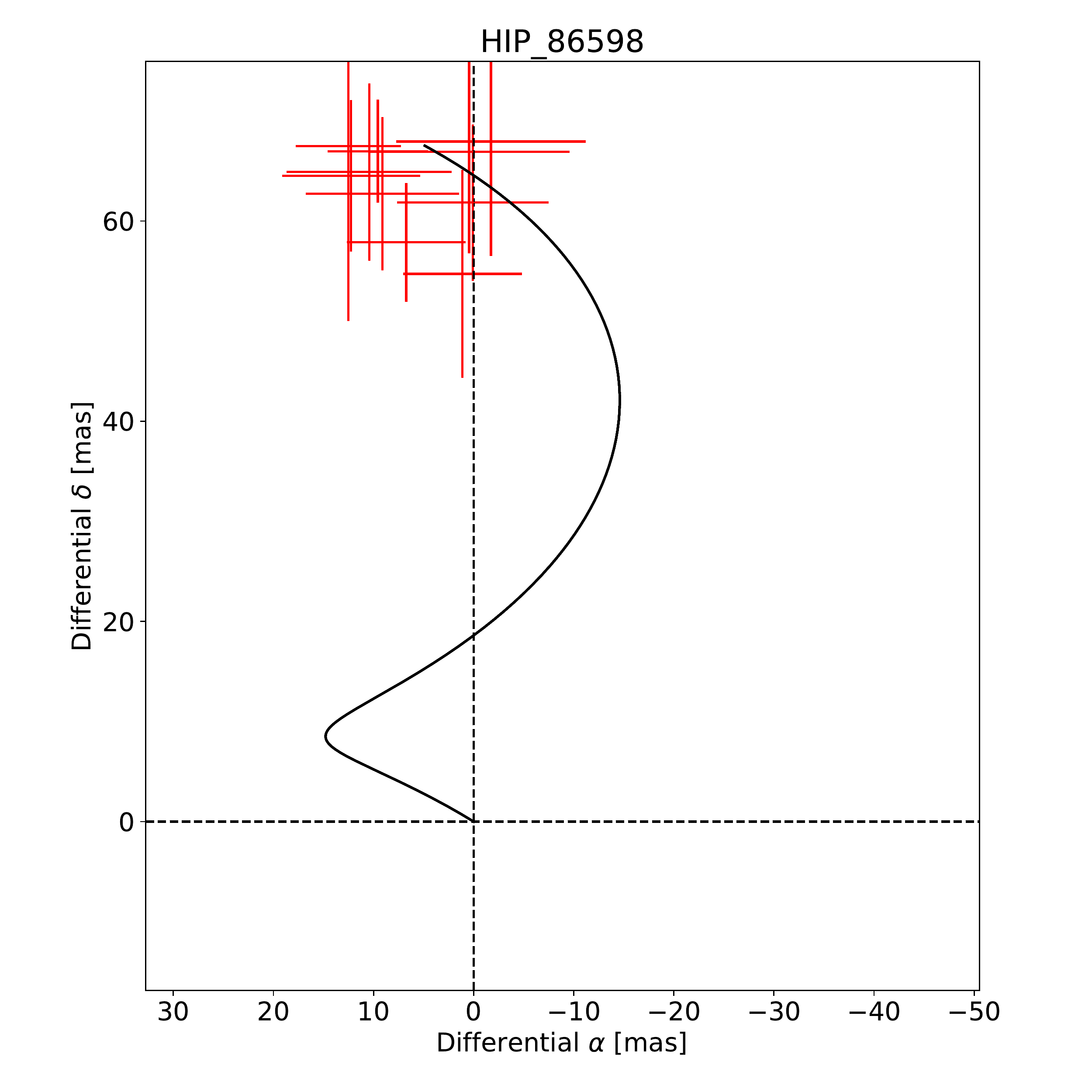}
\normalsize
\caption{Proper motion plot of point sources. Red crosses correspond to the RA/Dec motion of the point sources of the second epoch relative to the first epoch. The black line corresponds to the motion of a putative background star between the first and the second epochs. All motions of the point sources are compatible with the hypothesis of stationary background stars.}
\label{PM}
\end{figure}

The first-epoch observations revealed ten point sources in the field of view of IRDIS (Fig. \ref{irdis2015-2016}, left), none of them are in the field of view of the IFS image.
The second epoch confirms the detection of the previous ten point sources (Fig. \ref{irdis2015-2016}, right: white circles) and adds five new detections (Fig. \ref{irdis2015-2016}, right: black circles).\\

To determine the nature of these ten point sources (bound companions or stationary background stars), we \elodie{measured their astrometry}  in the TLOCI images and computed \elodie{their} motion \elodie{between the two epochs}.
The analysis of the proper motion plot  (Fig. \ref{PM}), which corresponds to the motion of each point source with respect to the proper motion of the star, allows us to conclude that \elodie{none} of the ten point sources are co-moving with the star, hence we can flag those as background stars.\\

Regarding the five new detections \elodie{from} 2016, we only have their H magnitude and one epoch of astrometry, which is not enough to perform a proper motion analysis or a colour-magnitude diagram (CMD). The CMD compares the colour of a candidate to the colours of template dwarfs, computed from their published spectra and SPHERE filters transmissions, to determine the probability of a point source being a planet or a background star, \elodie{ when a proper motion analysis is not possible}. The method is described in Appendix C of \citet{2018arXiv180700657B}. Candidates \#12, \#13, and \#14 are detected very close to the edge of the IRDIS image, with a projected separation up of $\sim$480\,au \elodie{assuming they are co-moving with the star}. Candidates \#11 \elodie{and \#15 are} very faint and \elodie{have} projected separations of $\sim$410\,au and $\sim$150\,au, \elodie{assuming common parallax with the star}. 
The status of these candidates is still undefined but based on their brightnesses and separations, \elodie{their probabilities} of being background stars are 100\%, 72.3\%, 100\%, 98.8\%, and 89\% for candidates \#11, \#12, \#13, \#14, and \#15, respectively, based on the Besançon Galaxy Model \citep{2003A&A...409..523R}. The angular separation and the position angle of all companion candidates are reported in Table \ref{tabcc}.\\

\begin{figure}[ht] 
\centering
\includegraphics[width=0.5\textwidth,clip]{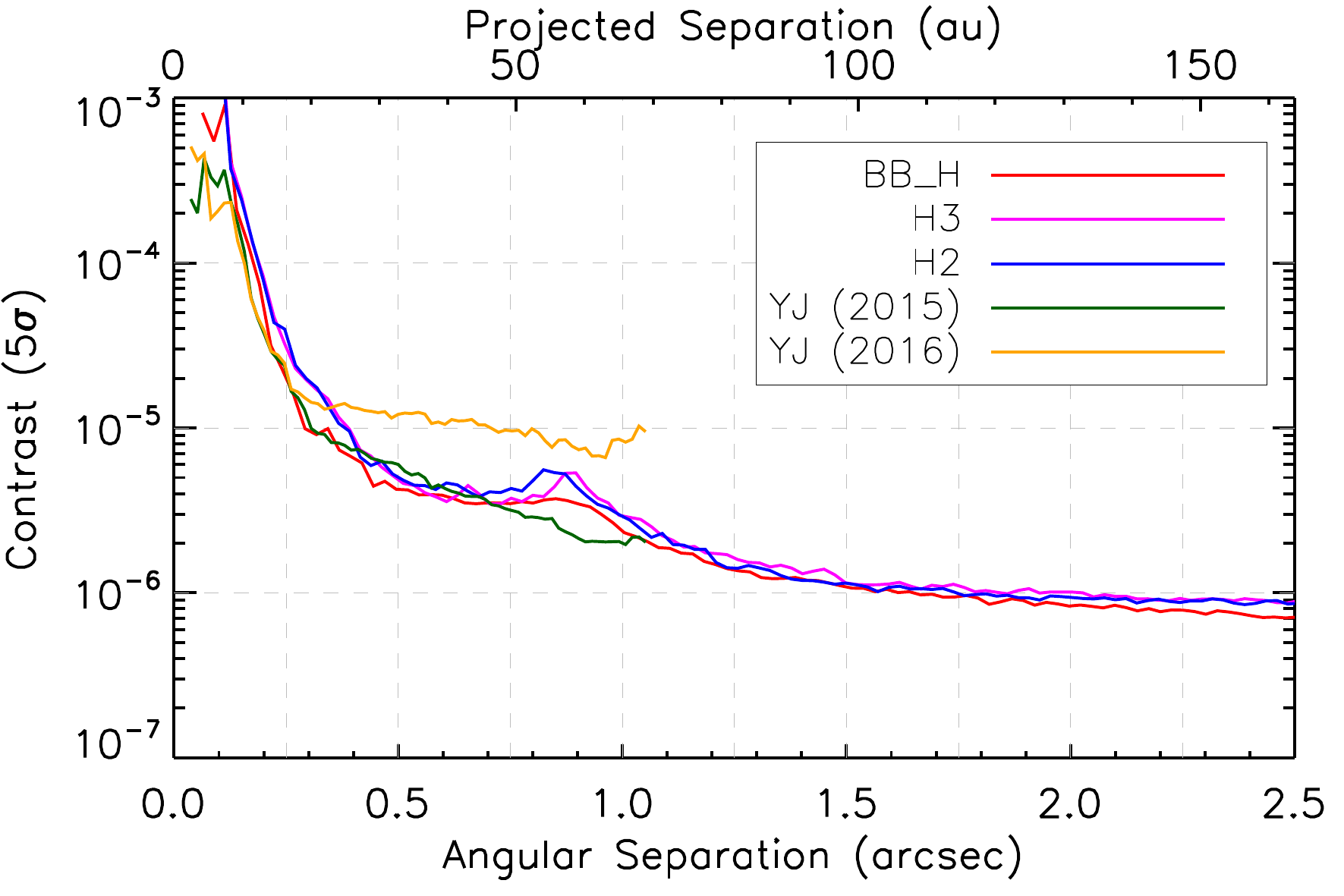}
\includegraphics[width=0.49\textwidth,clip]{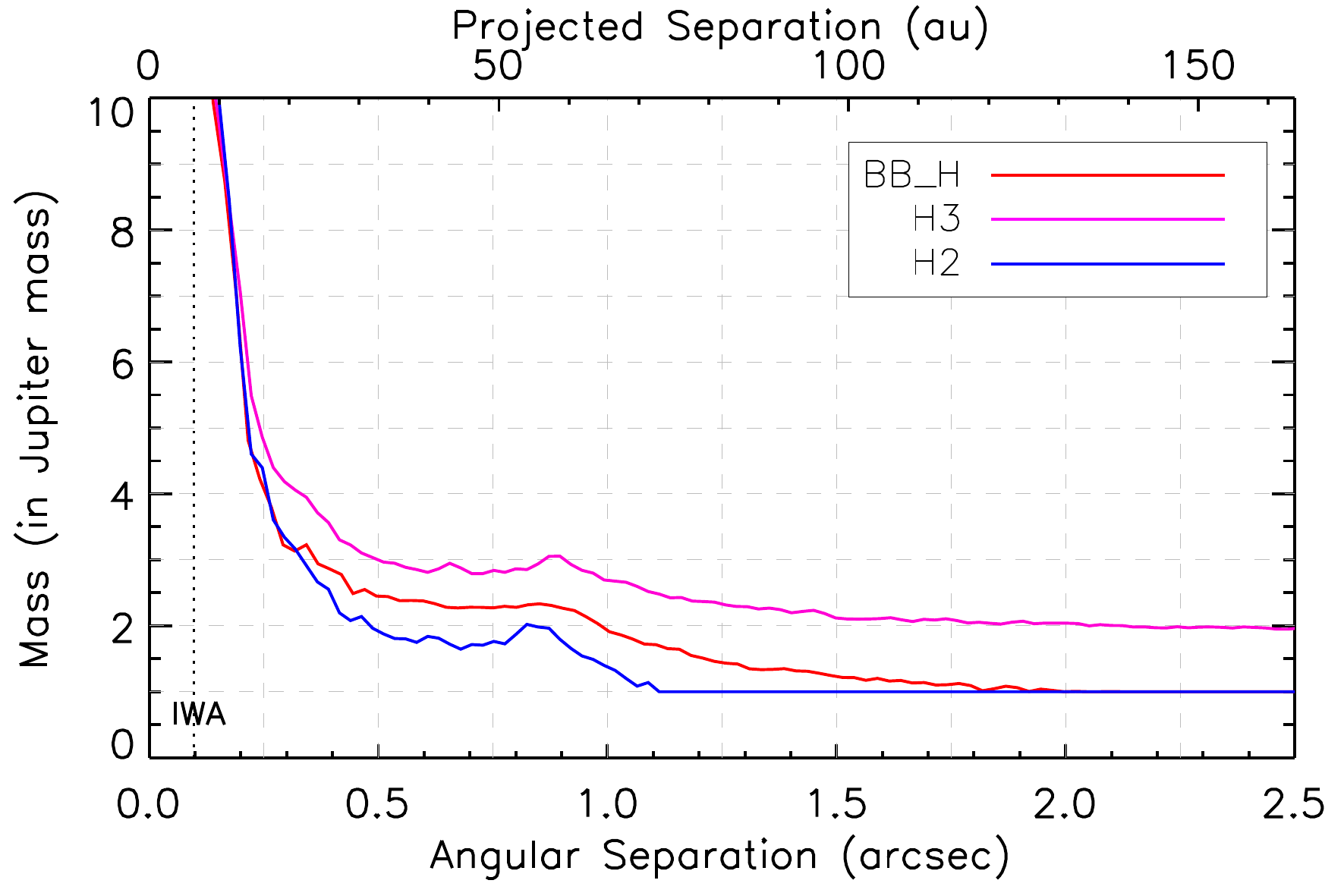}
\normalsize
\caption{Top: Contrast limits at 5$\sigma$ for the IRDIS observation in BB\_H (red), H3 (purple), H2 (blue) bands, and for the IFS observations in YJ bands (green for 2015 and yellow for 2016). These contrast limits are obtained with the TLOCI reduction.
Bottom: Detection limits in planet masses obtained with the BHAC-2015+COND evolutionary models \citep{2015A&A...577A..42B} for an age of 23\,$\pm$\,3\,Myr, for the IRDIS BB\_H (red), H2 (purple), and H3 (blue) bands. IWA correspond to the inner working angle of the coronagraph.}
\label{jupmass}
\end{figure}            

\begin{figure*}[h] 
\centering
\includegraphics[width=1.0\textwidth,clip]{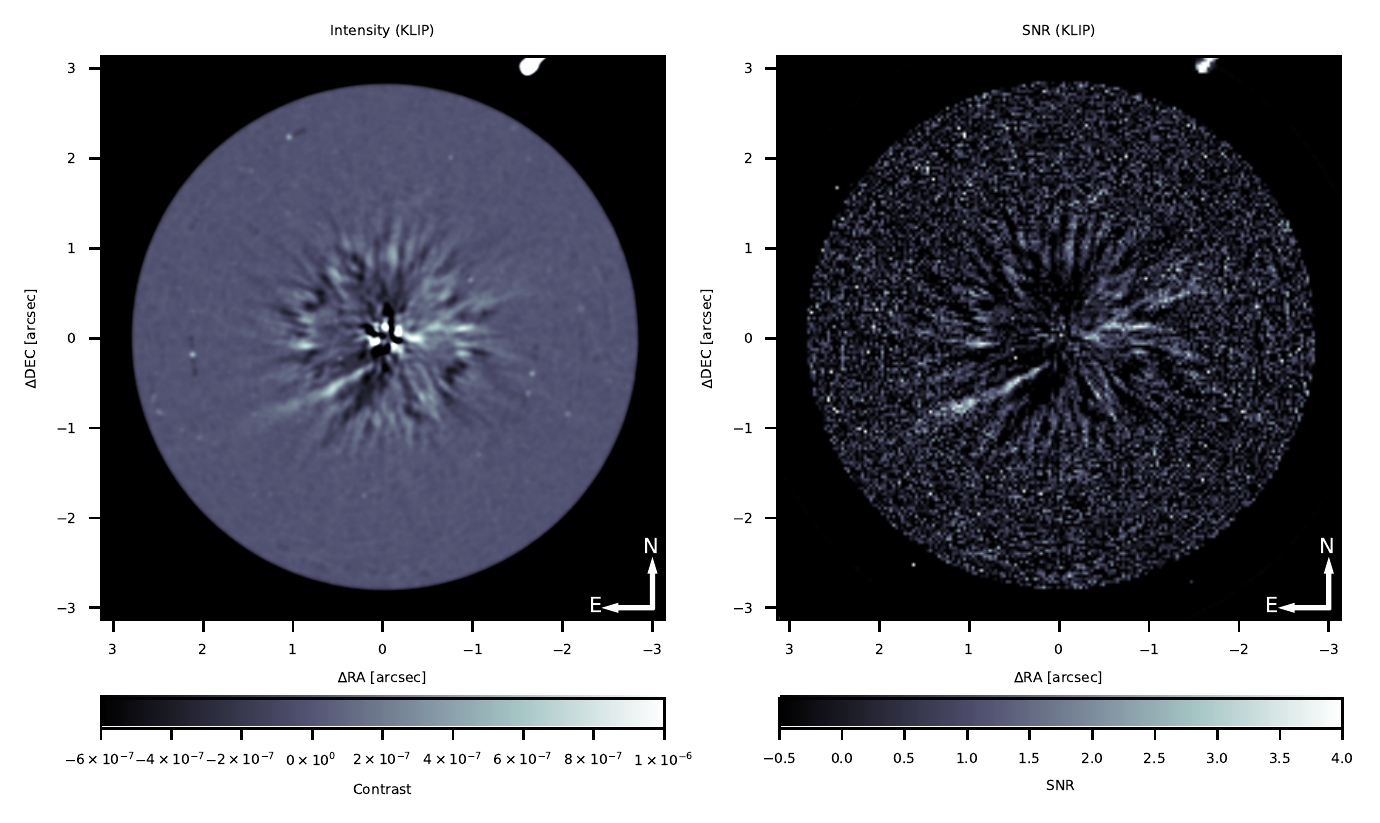}
\normalsize
\caption{Left: Zoom into the disc in the KLIP reduction of the May 2016 data. North is up, east is to the left. Right: Signal-to-noise map derived from the KLIP reduction (same orientation).}
\label{disksnr}
\end{figure*}

Using the \textit{SpeCal} pipeline \citep{2018A&A...615A..92G}, we derived the contrast limits in H2, H3, and BB\_H bands for IRDIS, and in YJ for IFS, \elodie{accounting} for TLOCI/ADI photometric biases due to self-subtraction \elodie{and for the coronagraph transmission}. \elodie{The contrast limits are presented} in the top panel of Fig. \ref{jupmass}. 
We converted the IRDIS contrast into detection limits expressed in Jupiter masses ($M_J$) using the BHAC-2015+COND evolutionary models \citep{2015A&A...577A..42B} assuming an age of 23\,$\pm$\,3\,Myr (isochrones are interpolated for these given ages). \elodie{These detection limits are presented in the} bottom panel of Fig. \ref{jupmass}.
The detection limits in Jupiter masses were not computed for the IFS data, because the conversion from contrast to planet mass is dependent on the spectrum assumed for the planets that we aim to detect.
The detection limits in BB\_H (H2) \elodie{rule out} the presence of companions more massive than 1\,$M_J$ beyond angular separations of 2\arcsec\,(1.1\arcsec), 2\,$M_J$ beyond angular separations of 1\arcsec\,(0.5\arcsec), and 4\,$M_J$ beyond angular separations of 0.25\arcsec\,(0.25\arcsec). \elodie{We} note that due to the high inclination of the system (see Sect. \ref{disk}), Jovian (resp. 2\,$M_J$ and 4\,$M_J$) companions at 80\,au (resp. 35\,au and 15\,au) located angularly close to the star at the time of observations cannot be excluded.

\section{Debris disc analysis}
\label{disk}

The first epoch of IRDIS data showed \elodie{the} weak detection of a disc seen at high inclination ($>$80\deg), with a position angle (PA) around 120\deg, at distances shorter than 1\arcsec\,from the star. Only the southeast side was detected in the first epoch. 
The detection of the disc was confirmed by the second epoch observations (Fig. \ref{disksnr}, left) in which we found the same recurrent pattern on the south-east side, while the north-west side was revealed.  \\

Figure \ref{disksnr} (right) shows the signal-to-noise ratio (S/N) map, per element of resolution, of the disc for the KLIP reduction. To obtain this S/N-map, we binned the KLIP image in such a way that one pixel corresponds to a resolution element. Then we divided the binned image by a radial map of the standard deviation of the KLIP image measured in  \elodie{annuli} (one resolution element width), which we use to estimate the noise in the image (the disc \elodie{was} not masked).
The disc is detected with a low S/N per resolution element, around 3-4 for the sout-east part and 1-2 for the north-west part. These low S/N can be explained by the nature of the noise (not Gaussian) at the position of the disc, which is dominated by large structures. These large structures are speckles \elodie{that} were not well suppressed by the \elodie{data} post-processing, \elodie{which} strongly affects the value of the noise at short separations and \elodie{at the edge} of the correction area of the adaptive optics ($\sim$1\arcsec). However, the south-east part of the disc is clearly different from the residual speckles pattern. We obtained an integrated S/N ratio of 16.8 for the south-east part. The integrated S/N ratio corresponds to the ratio between the integrated signal of the disc and the square root of the integrated square noise in the same area. This value represents the visual detectability of the disc and allows us to confirm the detection.
For the north-west part, the detection is very marginal but it is the most compelling explanation according to the disc analysis made in Sects. \ref{diskmodel}, \ref{secbestmodel}, and \ref{fluxratio}, assuming that the detected disc is a ring-like structure. We have chosen the assumption of a ring-like structure because the disc appears as a narrow curved feature with an offset from the central star. This observation is compatible with a ring-like structure or at least a belt, rather than an edge-on filled disc. However, other unusual configurations could be possible but cannot be confirmed or invalidated in this paper.\\

Assuming the ring-like structure of the disc, a brightness asymmetry in surface brightness is seen between the south-east part (the brightest) and the north-west part (barely detected). In the following, we will refer to this asymmetry as the two-sided asymmetry.
The disc is also asymmetrical on both sides of the major axis, the northern part being undetected. 
This "north-south" asymmetry suggests that the southern part of the disc is the closest to the earth, assuming forward scattering dominates, while the northern part would be the backside of the disc. 
In the following, this second asymmetry will be referred to as the "frontward" asymmetry. 
Despite the self-subtraction of the disc caused by KLIP/ADI, we observe an expected decreasing brightness profile from small to large scattering angles. In several cases, the brightness profile can have a strong peak of scattering efficiency at very small \elodie{scattering} angles \citep{2015ApJ...811...67H,2017A&A...599A.108M}.\\

\subsection{Disc model}
\label{diskmodel}
\begin{figure*}[ht] 
\centering
\includegraphics[width=1.0\textwidth,clip]{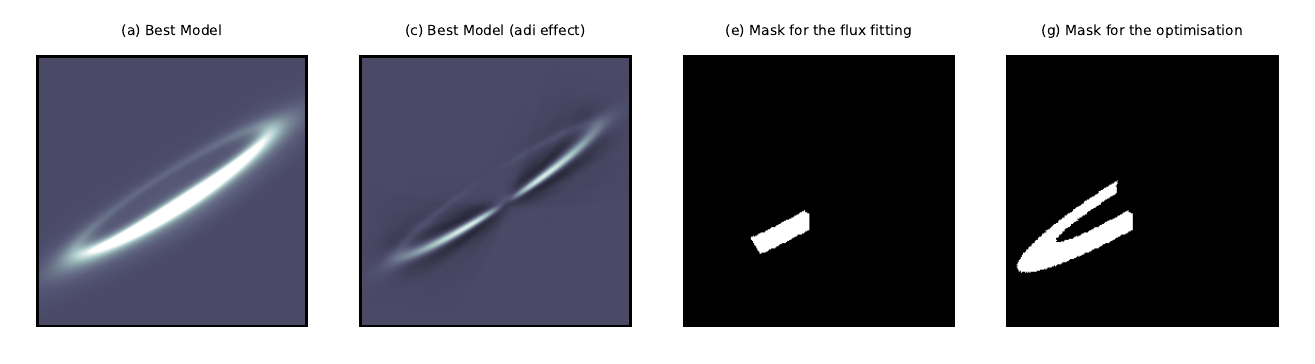}\\
\includegraphics[width=1.0\textwidth,clip]{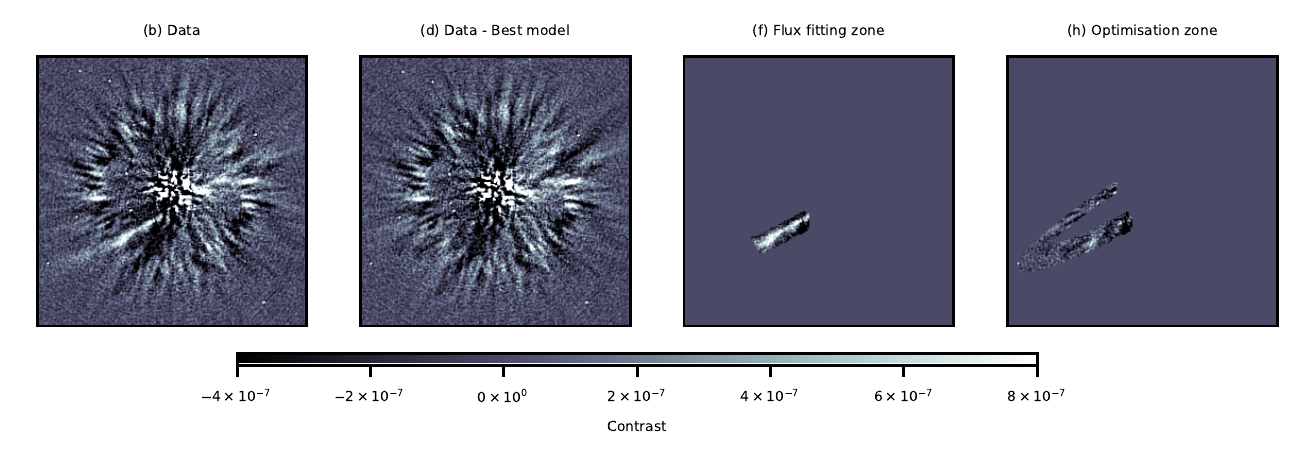}
\normalsize
\caption{Masks, data, and models used to determine the best model of the disc. 
a: Best model of the disc. 
b: KLIP-reduced data. 
c: Best model of the disc after KLIP/ADI reduction.
d: Data subtracted by the best model, after KLIP/ADI reduction. 
e: Mask used to select the part of the disc where the intensity scaling is done. 
f: Mask applied to the data for the intensity scaling of the disc model. 
g: Mask used to select the part of the disc where the $\chi^2$ minimization is done. 
h: Mask applied to the data, subtracted by the best model, for $\chi^2$ minimisation.
Masks e and g are based on approximate ellipse parameters, close to the disc parameters.} 
\label{bestmodel}
\end{figure*}

To derive the parameters of the disc, we used a model fitting method based on the \textit{GRaTer} \elodie{code} \citep{1999A&A...348..557A}. 
Given the low S/N of the disc, we could not directly measure the disc's scattering phase function. We used an analytical function to derive the disc's geometric parameters.
The model assumes a radial surface density distribution described by \citet{1999A&A...348..557A}, which decreases radially from $R_0$, the disc radius (in au), inwards and outwards as a power law with slopes $\alpha_{in}$ and $\alpha_{out}$. 
We used the grain scattering phase function defined by \citet{1941ApJ....93...70H}, parametrized by the coefficient of anisotropy of scattering $g$ (defined between -1 and 1, -1 meaning backward scattering and 1 meaning forward scattering).
We note $i$ the inclination of the disc (in degrees), $PA$  the position angle (in degrees), and $h$ the aspect ratio between the height of the disc and \elodie{its} radius $R_0$.
The simulated disc has no eccentricity, but it is possible to add two orthogonal offsets to simulate the effect of a small eccentricity of the disc (at first order).\\

\begin{table}[ht!]
\caption{Parameter used to generate the grid of models.}
\footnotesize
\centering
\begin{tabular}{ c c c}
\hline 
\hline
Parameters & Values$^{(1)}$ & Best model$^{(2)}$\\ 
\hline
$R_0$[au] & [81 ; 86 ; 90 ; 95 ; 99 ;& 86\\
          &  104 ; 109 ; 113 ; 118]&\\ 
$\alpha_{in}$ & [2 ; 5 ; 10 ; 20]& 10\\
$\alpha_{out}$ & [-2 ; -5 ; -10 ; -20]& -5\\
$i$[\deg] & [81 ; 82 ; 83 ; 84 ; 85] & 82\\
$PA$[\deg] & [120.5 ; 121 ; 121.5 ; 122 ;& 122.5\\
 & 122.5 ; 123 ; 123.5 ; 124]& \\
$g$ & [0.25 ; 0.5 ; 0.75]& 0.5\\
$h$ & [0.005 ; 0.01 ; 0.025 ; 0.05]& 0.005\\
\hline
\end{tabular}
\flushleft
\vspace{0.1cm}
\textbf{Notes.} $^{(1)}$Values of parameters to generate the grid of models. $^{(2)}$Values of the best model after a $\chi^2$ minimization.
\normalsize
\label{tabgrille}
\end{table}

\elodie{Our} method \elodie{to characterize} the disc is similar to the one used in \citet{2016A&A...590L...7P}. 
We generate a grid of models with different \elodie{parameter values}  (Table \ref{tabgrille}), for a total of 69\,120 models. 
\elodie{Several} steps of the optimization \elodie{process} are illustrated in Fig. \ref{bestmodel}, \elodie{comparing the best model to the KLIP-processed IRDIS data from the second epoch}. 
Each model is convolved with the stellar PSF \elodie{for comparison} with the data (Fig. \ref{bestmodel}a). 
\elodie{The models are then} projected onto the eigenvectors of the coronagraphic images to apply the same biases \citep[self-subtraction,][]{2012A&A...545A.111M} as \elodie{those affecting} the data (Fig. \ref{bestmodel}c), similar to the forward modelling approach used by \citet{2016ApJ...817L...2C}.
To avoid over-sampling in the optimization process, we convolved the data and the model with a Gaussian of three pixels, corresponding to one resolution element.

\begin{figure*}[ht] 
\centering
\includegraphics[width=0.99\textwidth,clip]{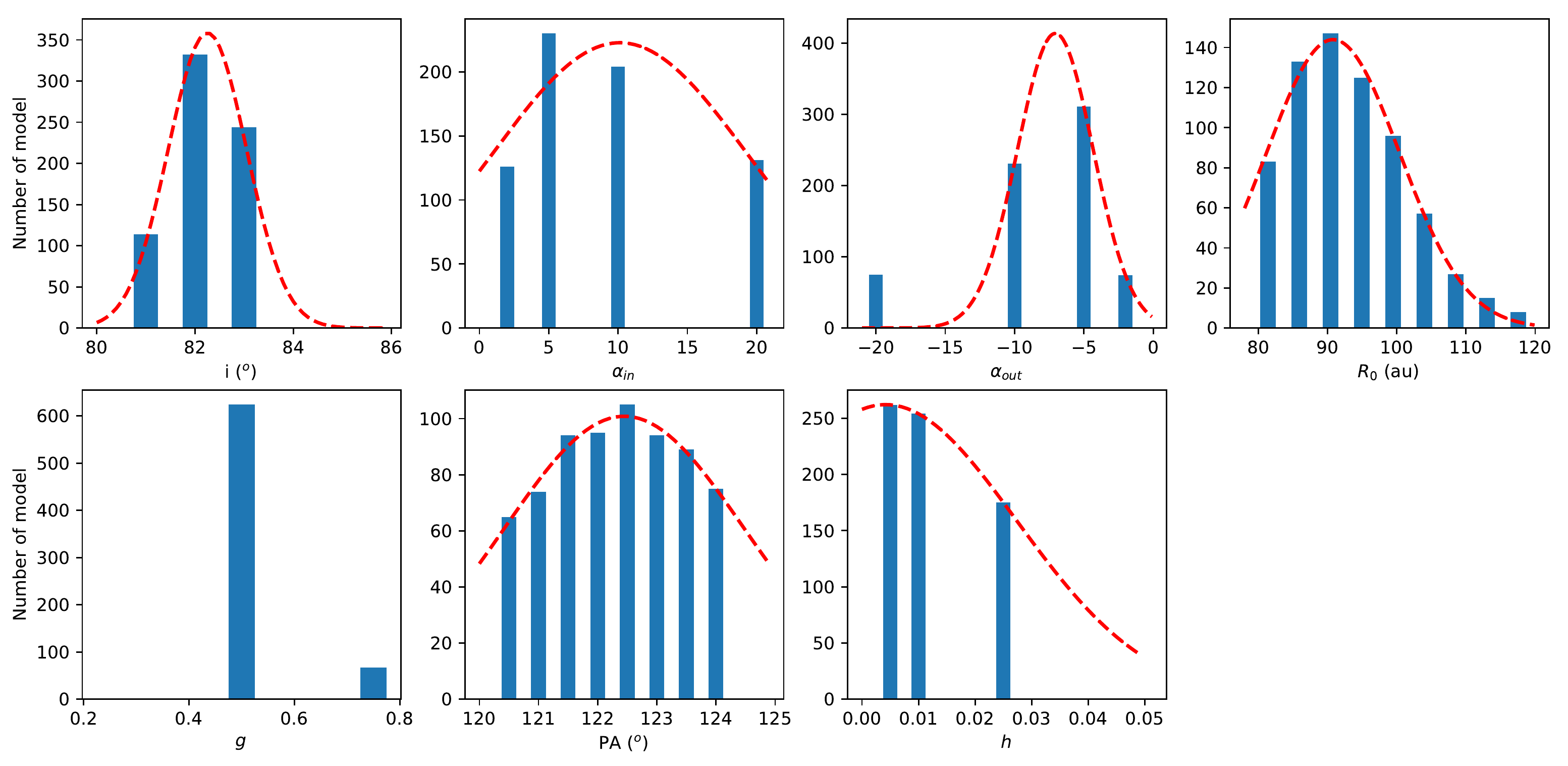}
\normalsize
\caption{Histogram of the 1\% best model for each parameter tested. The bars correspond to the values used for the model grid and presented in the Table \ref{tabgrille}.}
\label{histo1p}
\end{figure*}

Then, each model is subtracted from the data (Fig. \ref{bestmodel}d) with an intensity scaling factor to match the surface brightness of the model to the surface brightness of the disc (pixel to pixel). This intensity scaling factor is computed, using the $Amoeba$ function of IDL, on the brightest part of the disc only (the south-east part), using a mask (Fig. \ref{bestmodel}e) applied to the data (Fig. \ref{bestmodel}f).
Finally, for each model we estimate the goodness of fit ($\chi^2$) to determine which models are the closest to the observations.
The minimization is performed in the eastern part of the disc, with a specific mask (Fig. \ref{bestmodel}g) applied to the \elodie{residuals} between the model and the data (Fig. \ref{bestmodel}h). We used the eastern part of the disc instead of the full disc because the south-east to north-west surface brightness asymmetry of the disc cannot be reproduced with our symmetric debris disc model and because the northern part of the disc has a very low S/N.

\subsection{Best model}
\label{secbestmodel}
The parameter values corresponding to the best model are given in Table \ref{tabgrille}. The bright south-east part of the disc is well \elodie{fitted} by the best model. The north-west part of the disc is over-\elodie{fitted} by the model (\elodie{negative} pattern in Fig. \ref{bestmodel}d), which confirms the observed two-sided asymmetry in intensity. 

The frontward asymmetry, probably caused by forward scattering, is well reproduced \elodie{both} in the northern \elodie{and} in the southern part of the disc. For this model the best value found for $g$ is 0.5, but we note that the sampling of $g$ values \elodie{in our grid} is coarse.

To estimate the uncertainty on these values, we plot in Fig. \ref{histo1p} the histogram of the 1\% best models for each parameter (corresponding to 691 models). It is important to note that some of these 1\% best models are visually not very consistent with the disc image, but still have low $\chi^2$ \elodie{values}. This issue is mainly due to the low S/N of the disc and to the \elodie{poorly fitted two-sided} asymmetry. Therefore, these histograms provide a \elodie{trend} for \elodie{the goodness of fit of each} parameter value. In addition, we plot (as a red dashed line) the Gaussian fit of each histogram, except for the $g$ values due to the coarse sampling. Once again, because of the \elodie{low S/N} in many regions of the image, these fits should be taken with caution, especially for $\alpha_{in}$, $\alpha_{out}$, and $h$. The mean values and the standard deviation of the Gaussian fits are reported in Table \ref{tabhisto}.\\

\begin{figure}[ht] 
\centering
\includegraphics[width=0.45\textwidth,clip]{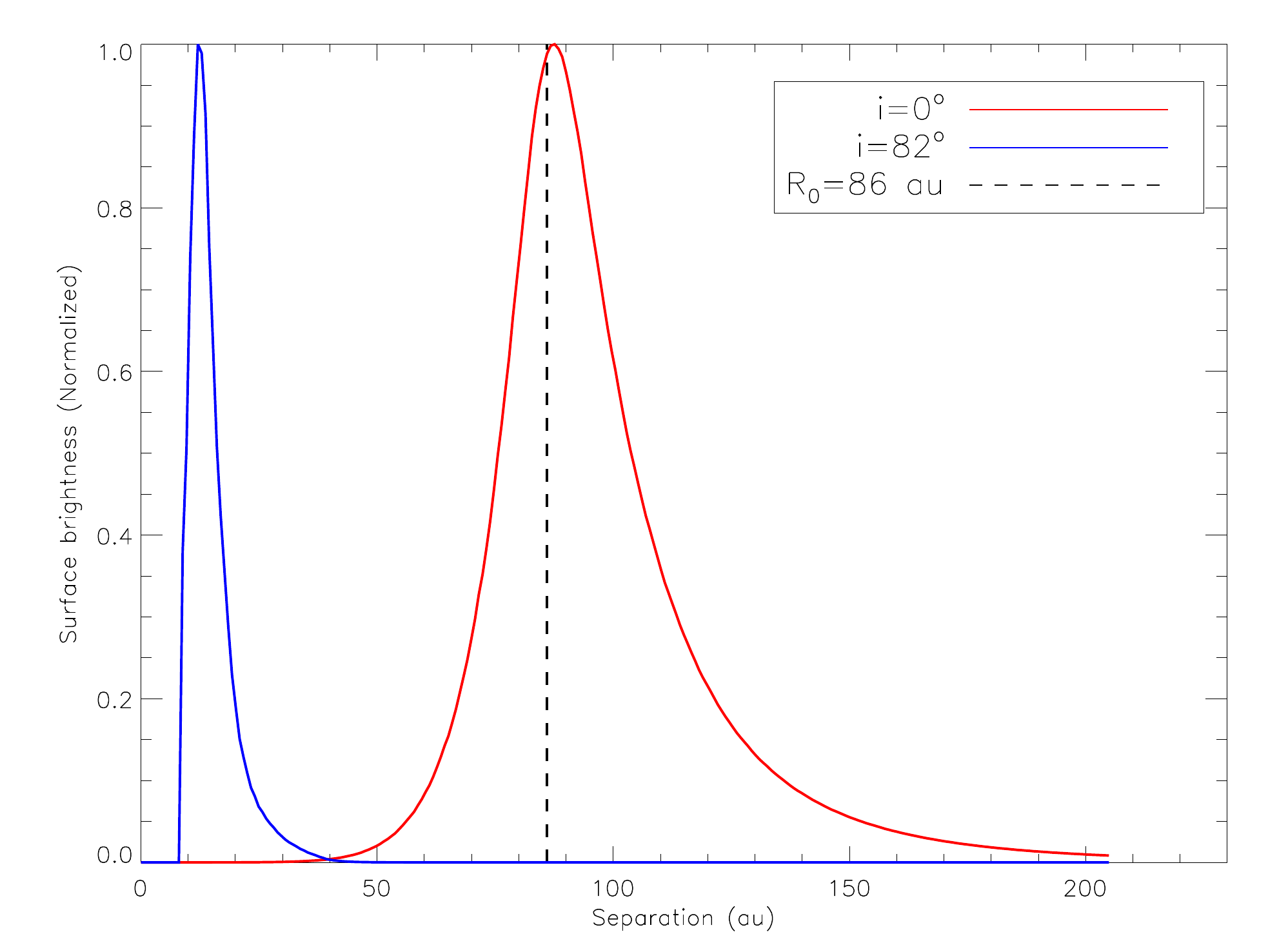}
\normalsize
\caption{Normalized radial profiles for the best model disc. Red line: Deprojected disc (face-on). Blue line: Viewing angle of 82\deg. The FWHM for the deprojected and inclined configurations are 27\,au and 6.4\,au, respectively. The black dashed line corresponds to the position of $R_0$.}\label{radialprof}
\end{figure}

The Gaussian fit for $\alpha_{out}$ favours values around -5 to -10. The situation is less clear for $\alpha_{in}$, for which the Gaussian fit has a large width, with a mean around 10. These results are not surprising because the S/N quickly becomes very poor beyond the main ring. However, these measurements are sufficient to confirm the ring-like nature of the disc, since the slope of the surface density gives a full width at half maximum (FWHM) around 27\,au (Fig. \ref{radialprof}).
The preferred values for the $g$ parameter are systematically higher than 0.5 but rarely higher than 0.75.
For the radius of the disc, $R_0$, the histogram is very well fitted by a Gaussian with a mean at $\sim$90\,au and a $\sigma_{R_0}\sim$10\,au. 
For the $h$ ratio, we can only estimate an upper limit of 0.025, but cannot discriminate between the thinner configurations 0.005 or 0.01, mainly because the $h$ ratio is constrained by the limited beam resolution. The value 0.025 is also present in 25\% of cases. This $h<0.025$ constraint is still compatible with the minimum "natural" scale height, $h\sim 0.04\pm0.02$, for debris discs in which small dust grain dynamics is controlled by radiation pressure \citep{2009A&A...505.1269T}.
The inclination $i$, which is 82\deg\,for the best model \elodie{of the grid}, is  estimated to 82.4$\pm$0.8\deg\,with the Gaussian fit.
Finally, the position angle $PA$ is constrained with a mean at 122.4\deg\,and a dispersion $\sigma_{PA}$\,=\,2\deg. This result is still in agreement with the best model \elodie{of the grid} (122.5\deg) and should be better constrained with new deeper observations. 
Using the values obtained for $\alpha_{in}$ and $\alpha_{out}$ with the best model, we determined the radial profile of the disc (Fig. \ref{radialprof}). 
If seen face-on, the disc would have a FWHM of 27\,au (red line). With an inclination of 82\deg\,the front of the disc (along the semi-minor axis) would have a FWHM of 6.4\,au (blue line).\\ 

\begin{table}[ht!]
\centering
\caption{Parameter values obtained with the 1\% best models.}
\footnotesize
\begin{tabular}{ c c}
\hline 
Parameters & Values \\ \hline
$R_0$ [au] & 90.8$\pm$9.6\\
$\alpha_{out}$ & -7.1$\pm$2.7\\
$\alpha_{in}$ & 10.1$\pm$9.3\\ 
$\Delta R$ [au] & 27\\
$i$ [\deg] & 82.3$\pm$0.8\\
$PA$ [\deg] & 122.4$\pm$2.0\\
$g$ & $\sim$0.5\\
$h$ & <0.025
\end{tabular}
\justify
\textbf{Notes.} The parameter $\Delta R$ corresponds to the FWHM of the ring, derived from the $\alpha_{in}$ and $\alpha_{out}$ parameters of the best model. The $R_0$, $\alpha_{in}$, $\alpha_{out}$, $i,$ and $PA$ values and errors are determined by fitting a Gaussian to the histogram (red dashed line). Due to the low number of points for $\alpha_{out}$ and $\alpha_{in}$, these values should be taken with caution. Moreover, $g$ and $h$ do not have enough points to perform a Gaussian fitting.
\normalsize
\label{tabhisto}
\end{table}

\subsection{South-east to north-west surface brightness ratio}
\label{fluxratio}

\begin{figure}[!ht] 
\centering
\includegraphics[width=0.45\textwidth,clip]{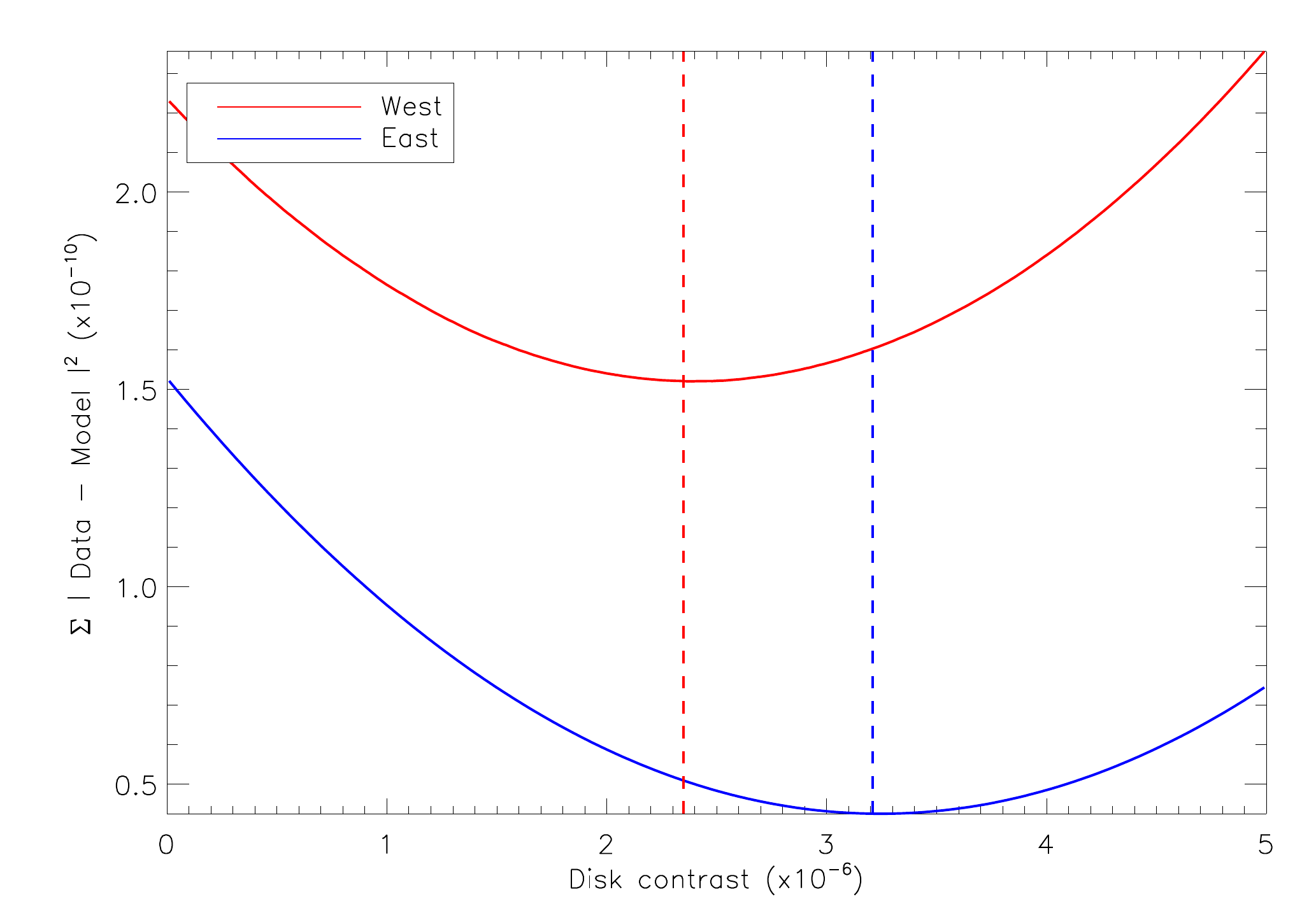}
\normalsize
\caption{Optimization criteria versus the surface brightness factor of the model for the south-east (blue) and north-west (red) part of the disc. Dashed lines correspond to the abscissa position of the minimum.}
\label{ratio}
\end{figure}

We measure the surface brightness ratio between the south-east and north-west parts ($\frac{\Phi_E}{\Phi_W}$) to \elodie{quantify} the two-sided asymmetry.
To do so, we used the best model derived from the previous analysis, \elodie{normalized to different surface brightness levels},  then subtracted \elodie{from} the raw data before the post-processing (negative injection of a fake disc in raw data). We \elodie{used} surface brightness factors \elodie{ranging from}  0 \elodie{to} $5.10^{-6}$ \elodie{with} steps of $10^{-8}$ in terms of contrast. With this analysis, we can determine \elodie{the} surface brightness \elodie{of} the south-east and \elodie{of the} north-west sides independently and compute the surface brightness ratio between \elodie{them}.\\

More precisely, for each tested surface brightness value we inject  a negative model in each raw frame at the same \elodie{PA} as the real disc (taking into account the field rotation), then we apply the KLIP post-processing algorithm with a truncation at ten modes. Therefore, the disc is more or less removed in the final obtained image, \elodie{depending on} the value of the injected surface brightness. 
To determine the optimal surface brightness factor, we minimize the mean squared error ($\Sigma$) in the residuals inside two specific masks, one for each side \elodie{of the disc}.

In Fig.~\ref{ratio}, we plot this criterion versus the surface brightness factor for both sides. The minimum of each plot gives the optimal surface brightness factor for the south-east and for the north-west side. The surface brightness ratio is then the ratio between these two coefficients. 
We find a value of $\frac{\Phi_E}{\Phi_W}=0.73\pm0.18$. The uncertainty is given by a combination of the error ($3.5\times 10^{-7}$) due to the noise inside the correction area, which extends from 250\,mas to 1\arcsec\,and corresponds to the zone where the disc is located, and the systematic error ($10^{-8}$) for the method, which corresponds to the step of the \elodie{evaluated} surface brightness \elodie{factors}. Of course the dominant contribution is due to the error caused by the residual speckle subtraction in the correction area.\\

With this analysis, we are also able to determine the flux ratio, in scattered light, between the star and the disc $\frac{\Phi_{disc}}{\Phi_{\star}}$. To do so, we integrated the intensity of an off-axis image of the star (which is not saturated) and the surface brightness of the best model, which is adjusted in intensity to correspond to the real disc. We find $\frac{\Phi_{disc}}{\Phi_{\star}}=1.4\pm0.22\times10^{-4}$. We note that this value is relatively close to the one derived for the thermal emission fractional luminosity by \citet{2016ApJ...826..123M}, $1.2\pm0.4\times10^{-4}$.

\section{Discussion}
\label{disc}

The disc presents two kinds of asymmetries: the frontward asymmetry, between north and south, and the two-sided asymmetry, between south-east and north-west. 
The first one can easily be explained by a forward scattering effect of the dust, making the front side of the disc brighter than its back side \citep{2018ARA&A..56..541H}. In this case, the effect of scattering can be modelled, in the same way as in Sect. \ref{disk}, where we found a value of the g-factor compatible with the forward scattering effect.
As for the south-west to north-east asymmetry, corresponding to an azimuthal asymmetry in the debris ring, there are several possible explanations, which can be divided into two categories: scenarios assuming there is physically more \elodie{dust} in the brighter regions, and those assuming that these regions are bright because they are closer to the star.\\

If the south-west side corresponds to a physical local over-density, a \elodie{first} explanation is the perturbations of a massive planet, which \elodie{can} create azimuthal inhomogeneities at the location of mean motion resonances \citep[e.g. ][]{2003ApJ...588.1110K, 2008A&A...480..551R, 2012A&A...547A..92T}. 
The presence of such a giant planet sculpting the main ring is not detected \elodie{in our data} after subtraction \elodie{of the best disc model}  (Fig. \ref{bestmodel}d). If there is a planet inside the disc, its mass is below our detection limits (1-2 M$_J$) or the planet is hidden by the residuals close to the star. However, super-earth mass planets, \elodie{which are below our detection limits (see Sect. \ref{cc})}, can also create \elodie{such disc} asymmetries \citep{2016ApJ...827..125L}, \elodie{and this explanation for the asymmetry thus cannot be ruled out}.
The planetary hypothesis could be compounded by the fact that there seems to be a sharp drop in surface brightness beyond the main ring, corresponding to a surface density decreasing with a slope $-10\lesssim\alpha_{out}\lesssim-5$ that is much steeper than the canonical -1.5 slope expected for a small grain halo beyond an unperturbed ring \citep{2006ApJ...648..652S, 2008A&A...481..713T}. However, because of the very poor S/N beyond the main ring, the $\alpha_{out}$ value is only constrained over a narrow radial domain beyond $R_{0}$, for which departures from a -1.5 slope could in fact be obtained even for an unperturbed system (the asymptotical -1.5 slope being reached further out, see \citealt{2012A&A...547A..92T}).\\

\elodie{A second explanation for an} azimuthal over-density is the result of the recent breakup of a massive planetesimal, which releases dusty fragments with highly eccentric orbits all passing through the initial breakup location. This can create an observable long-lived two-sided asymmetry \citep{2014MNRAS.440.3757J, 2015A&A...573A..39K}. However, the fainter side (opposite to the breakup location) should in this case be more radially extended than the bright side, whereas we observe the opposite, even though this trend should be taken with caution because the luminosity profile beyond the main ring is poorly constrained in the present images.\\

An alternative scenario is that the south-west to north-east asymmetry is not due to azimuthal inhomogeneities but to the fact that the disc is eccentric, with its periastron located somewhere in the western side, scattering more stellar light because it is closer to the star. This "pericenter glow" effect has been identified by \citet{1999ApJ...527..918W} and is suspected to be witnessed in several discs. More generally, an eccentric ring of parent bodies, coupled to the size-dependent effect of radiation pressure, can create strong brightness asymmetries. The thorough exploration by \citet{2016ApJ...827..125L} showed that, for some disc geometries and viewing angles, asymmetries resembling the one observed here can be obtained (see for example Fig.~9 of that paper). Of course, the eccentric ring scenario also requires an additional force (perturbations by a planetary object or a companion star) to explain this eccentricity. However, as with the giant-breakup scenario, the fainter side of the disc should be radially more extended than the bright side, in apparent contradiction with the present image.\\

At the present time, the large uncertainties \elodie{caused by} the \elodie{low} S/N of our observations do not allow us to easily discriminate between these different scenarios. Future and deeper observations, in particular to constrain the surface brightness beyond the main ring, are clearly needed.\\

We also report the apparent discrepancy between the disc's radial location inferred here, $\sim86$\,au, and the one derived by \citet{2016ApJ...826..123M}, $58\pm13$\,au, by fitting the system's SED. However, this difference can be attributed to the fact that the SED fitting was done assuming black-body temperatures for the grains. As rightfully underlined by \citet{2016ApJ...826..123M} themselves, smaller grains, which are the ones likely to dominate the thermal emission up to mid-IR wavelengths, are ineffective emitters that can be much hotter than black bodies at the same radial distance. As such, the 58\,au value should be regarded as a lower limit that could easily be increased by a factor of two with more realistic modelling \citep{2016ApJ...826..123M}, making it compatible with our present measurement.
In order to correct this effect, we used the correction factor described in \citet{2015MNRAS.454.3207P} (Eqs. 8 and 9), and we corrected the dust radius and obtained several values, depending on the dust composition (we need more measurements on the SED to discriminate the dust composition). We found a dust radius of 203.3$\pm$74.0\,au (50\% astrosilicate, 50\% vaccum), 195.7$\pm$65.8\,au (50\% astrosilicate, 50\% ice), 274.2$\pm$110.5\,au (100\% astrosilicate), 226,7$\pm$84.7\,au (50\% astrosilicate, 50\% carbon), and 228.8$\pm$82.7\,au (100\% carbon). 
However, the radii found with the correction of the SED are more extended than the radius measured with SPHERE. This difference could be explained by the lack of measurements in the SED or because the disc is more extended than what we observe with SPHERE, but unseen or suppressed by the ADI algorithm (which acts as a low spatial frequency filter).
Our assumptions for these computations are: the bolometric luminosity of the star $L/L_{\sun}$=1.52 \citep{2015MNRAS.454.3207P}, the dust temperature $T_{dust}$=43$\pm$5 \citep{2016ApJ...826..123M}, and the coefficients A and B of the $\Gamma$ ratio from \citet[Table 4,][]{2015MNRAS.454.3207P}.

\section{Conclusion}
\label{concl}

The debris disc around the F9V star HD\,160305 was detected and resolved for the first time in scattered light with SPHERE at two different epochs. We also detected several point sources around the central star.
In this paper we show that ten of these point sources are background stars, while the last five, only detected in the second epoch, can probably be classified as contaminant background stars as well.

We are able to derive important constraints on the disc's morphology. It has a ring-like structure, and is very inclined ($\sim$82\deg) with respect to the line of sight, with a radius of approximately 86-90\,au and an estimated deprojected width of 27\,au. We also determine that the southern part of the disc is probably the front side of the ring, appearing much brighter than the northern part due to the forward scattering effect of the dust.

We also detected a two-sided surface brightness asymmetry, between the south-east and the north-west parts. The surface brightness ratio between the two parts is 0.73$\pm$0.18 and the fractional luminosity of the disc relative to the star is around $1.4\times10^{-4}$.
Several scenarios could explain this asymmetry, \elodie{such as} the interaction with a massive companion, a recent and massive collision of planetesimals, or a pericenter glow effect due to a potential eccentricity of the disc. However, due to the faintness of the disc in our current data we cannot discriminate between these different scenarios without follow-up observations.

\begin{acknowledgements}
SPHERE is an instrument designed and built by a consortium consisting of IPAG (Grenoble, France), MPIA (Heidelberg, Germany), LAM (Marseille, France), LESIA (Paris, France), Laboratoire Lagrange (Nice, France), INAF–Osservatorio di Padova (Italy), Observatoire de Genève (Switzerland), ETH Zurich (Switzerland), NOVA (Netherlands), ONERA (France) and ASTRON (Netherlands) in collaboration with ESO. SPHERE was funded by ESO, with additional contributions from CNRS (France), MPIA (Germany), INAF (Italy), FINES (Switzerland), and NOVA (Netherlands).  SPHERE also received funding from the European Commission Sixth and Seventh Framework Programmes as part of the Optical Infrared Coordination Network for Astronomy (OPTICON) under grant number RII3-Ct-2004-001566 for FP6 (2004–2008), grant number 226604 for FP7 (2009–2012), and grant number 312430 for FP7 (2013–2016). We also acknowledge financial support from the Programme National de Planétologie (PNP) and the Programme National de Physique Stellaire (PNPS) of CNRS-INSU in France. This work has also been supported by a grant from the French Labex OSUG@2020 (Investissements d’avenir – ANR10 LABX56). The project is supported by CNRS, by the Agence Nationale de la Recherche (ANR-14-CE33-0018). It has also been carried out within the frame of the National Centre for Competence in Research PlanetS supported by the Swiss National Science Foundation (SNSF). MRM, HMS, and SD are pleased  to acknowledge this financial support of the SNSF. Finally, this work has made use of the the SPHERE Data Centre, jointly operated by OSUG/IPAG (Grenoble), PYTHEAS/LAM/CESAM (Marseille), OCA/Lagrange (Nice) and Observtoire de Paris/LESIA (Paris). We thank P. Delorme and E. Lagadec (SPHERE Data Centre) for their efficient help during the data reduction process.
R.\,G. and S.\,D. acknowledge support from the "Progetti Premiali" funding scheme of the Italian Ministry of Education, University, and Research. This work has been supported by the project PRIN INAF 2016 The Cradle of Life - GENESIS-SKA (General Conditions in Early Planetary Systems for the rise of life with SKA).
J.\,O. and C.\,P. acknowledge financial support from the ICM (Iniciativa Cient\'ifica Milenio) via the N\'ucleo Milenio de Formaci\'on Planetaria grant, from the Universidad de Valpara\'iso. 
J.\,O. also acknowledge financial support from Fondecyt (grant 1180395).
C.\,P. also acknowledge financial support from Fondecyt (grant 3190691).
F.\,Me. acknowledges funding from ANR of France under contract number ANR-16-CE31-0013.
D.\,M. acknowledges support from the ESO-Government of Chile Joint Comittee program 'Direct imaging and characterization of exoplanets'.
A.-L.\,M. acknowledges financial support from the F.R.S.-FNRS.
\end{acknowledgements}

\bibliography{HIP86598_disk}

\clearpage
\appendix
\section{Derivation of the stellar rotation axis}
\label{sini}
To derive the inclination of the stellar rotation axis, $i_\star$, we used the formula
\begin{equation}
sin(i_\star) = \frac{P \times v_{sini}}{50.576 \times R}
,\end{equation}
where P is the period of the stellar rotation (in days), v$_{sini}$ the projected rotational velocity (in km.s$^{-1}$), R the stellar radius (in solar radii), and 50.576 a constant to transform units from centimetre and s to solar radii and days.
For HD\,160305, we assume P\,=\,1.336\,$\pm$\,0.008 days \citep{2017A&A...600A..83M}, v$_{sini}$\,=\,37\,km.s$^{-1}$ (as no error bars were reported, we assume 10\% of uncertainties in the $v_{sini}$ value).
We derived a stellar luminosity L\,=\,1.6\,$\pm$\,0.2\,L$_{\sun}$ and a stellar radius R\,=\,1.15\,$\pm$\,0.18\,R$_{\sun}$ using the new Gaia distance from the DR2 \citep[65.51\,$\pm$\,0.23\,pc,][]{2018A&A...616A...1G}, the magnitude V\,=\,8.405 \citep{2012AcA....62...67K}, an effective temperature T\,=\,6065\,K \citep{2013ApJS..208....9P}, and a bolometric correction BC$_V$\,=\,-0.05 magnitude \citep{2013ApJS..208....9P} .
Finally, we can derive a stellar rotation axis of $i_\star$\,=\,58\deg$^{+18}_{-10}$.
This value is different from the disc inclination found (~82\deg) and could suggest a misalignment between the stellar equatorial plane and the disc inclination. 
It is possible that the asymmetry observed in the disc was a consequence of this misalignment. However, in this paper we have no proof for this hypothesis, which could be investigated in a future work.

\section{Additional data}

\begin{table*}[]
\centering
\footnotesize
\caption{Setup of observations.} 
\begin{tabular}{ c c c c c c c c c c c }
\hline 
Date & Mode & Filter & DIT & NDIT & T$_{exp}$ & Rotation & Seeing & $\tau_0$ & TN & pixscale\\
- & - & - & (s) & - & (s) & (\deg) & (\arcsec) & (ms) & (\deg) & (mas)\\ \hline
2015-05-13$^{(a)}$ & IRDIS & H2\,+\,H3 & 64 & 64 & 4096 & 34.58 & 0.7-1.0 & 3-4 & -1.712\,$\pm$\,0.063 & 12.242\,$\pm$\,0.062\\
2015-05-13$^{(a)}$ & IFS & YJ & 64 & 64 & 4096 & 34.58 & 0.7-1.0 & 3-4 & -1.712\,$\pm$\,0.063 & 7.46$\,\pm$\,0.02\\
2016-05-23$^{(b)}$ & IRDIS & BB\_H & 32 & 144 & 4608 & 40.20 & 0.5-0.7 & 3-6 & -1.675\,$\pm$\,0.080 & 12.247\,$\pm$\,0.017\\
2016-05-23$^{(b)}$ & IFS & YJ & 32 & 144 & 4608 & 40.20 & 0.5-0.7 & 3-6 &-1.675\,$\pm$\,0.080 & 7.46$\,\pm$\,0.02\\
\end{tabular}
\justify
\textbf{Notes.} Column 1 \textendash\, Date of the observation (\textit{$^{(a)}$} is programme 95.C-0298(A), \textit{$^{(b)}$} is programme 97.C-0865). Column 2 \textendash\, Observation mode (IRDIS or IFS). Column 3 \textendash\, Filter used. Column 4 \textendash\, Individual time for each individual image. Column 5 \textendash\, Total number of images. Column 6 \textendash\, Total time of the observation. Column 7 \textendash\, Total field rotation of the observation. Column 8 \textendash\, Seeing measured by the telescope. Column 9 \textendash\, Coherence time measured at the telescope. Column 10 \textendash\, Value of the true-north orientation, and Column 11 \textendash\, Angular size of a pixel \citep{2016SPIE.9908E..34M}.
\normalsize
\label{logobs}
\end{table*}

\begin{table*}[]
\centering
\footnotesize
\caption{Photometric and astrometric measurements of point sources detected in the first epoch (columns 1 to 5) and the second epoch (columns 6 to 8).}
\begin{tabular}{ c c c c c c c c c c }
\hline 
\# & $m_{H2}$ & $m_{H3}$ & Separation & PA & $m_{H}$ & Separation & PA & Probability & Status \\ 
- & - & - & (mas) & (\deg) & - & ($\arcsec$) & (\deg) & \% & - \\ \hline
1 & 16.77\,$\pm$\,0.08 & 16.78\,$\pm$\,0.08 & 3338.1\,$\pm$\,17.0 & 331.33\,$\pm$\,\,0.07 & 16.50\,$\pm$\,0.15 & 3392.5\,$\pm$\,4.9 & 332.01\,$\pm$\,0.09 & 13.5 & BS\\
2 & 20.40\,$\pm$\,0.08 & 20.28\,$\pm$\,0.08 & 3451.6\,$\pm$\,17.6 & 3.43\,$\pm$\,\,0.39 & 20.11\,$\pm$\,0.15 & 3519.7\,$\pm$\,5.1 & 3.57\,$\pm$\,0.28 & 71.5 & BS\\
3 & 22.69\,$\pm$\,0.19 & 22.56\,$\pm$\,0.15 & 3694.1\,$\pm$\,20.5 & 115.71\,$\pm$\,\,0.12 & 22.29\,$\pm$\,0.18 & 3677.7\,$\pm$\,7.7 & 114.72\,$\pm$\,0.10 & 93.7 & BS\\
4 & 19.08\,$\pm$\,0.08 & 19.02\,$\pm$\,0.08 & 4091.7\,$\pm$\,20.8 & 186.52\,$\pm$\,\,0.16 & 18.79\,$\pm$\,0.15 & 4037.2\,$\pm$\,5.8 & 186.60\,$\pm$\,0.15 & 57.9 & BS\\
5 & 22.72\,$\pm$\,0.16 & 22.57\,$\pm$\,0.21 & 4143.9\,$\pm$\,22.7 & 327.61\,$\pm$\,\,0.17 & 22.31\,$\pm$\,0.21 & 4193.3\,$\pm$\,7.8 & 328.20\,$\pm$\,0.13 & 97.4 & BS\\
6 & 17.53\,$\pm$\,0.08 & 17.58\,$\pm$\,0.08 & 4154.8\,$\pm$\,21.1 & 58.29\,$\pm$\,\,0.07 & 17.27\,$\pm$\,0.15 & 4191.2\,$\pm$\,5.9 & 57.66\,$\pm$\,0.08 & 30.9 & BS\\
7 & 21.74\,$\pm$\,0.10 & 21.74\,$\pm$\,0.10 & 5046.4\,$\pm$\,25.8 & 150.21\,$\pm$\,\,0.09 & 21.42\,$\pm$\,0.17 & 4992.9\,$\pm$\,7.5 & 149.86\,$\pm$\,0.09 & 98.1 & BS\\
8 & 19.93\,$\pm$\,0.08 & 19.93\,$\pm$\,0.08 & 5376.7\,$\pm$\,27.3 & 60.42\,$\pm$\,\,0.07 & 19.64\,$\pm$\,0.15 & 5415.8\,$\pm$\,7.7 & 59.89\,$\pm$\,0.08 & 90.0 & BS\\
9 & 21.98\,$\pm$\,0.13 & 21.93\,$\pm$\,0.14 & 5616.8\,$\pm$\,28.9 & 201.19\,$\pm$\,\,0.12 & 22.27\,$\pm$\,0.19 & 5554.2\,$\pm$\,9.0 & 201.46\,$\pm$\,0.12 & 99.8 & BS\\
10 & 21.20\,$\pm$\,0.19 & 20.71\,$\pm$\,0.10 & 7009.0\,$\pm$\,36.0 & 110.91\,$\pm$\,\,0.07 & 20.67\,$\pm$\,0.16 & 6985.8\,$\pm$\,10.2 & 110.40\,$\pm$\,0.08 & 99.6 & BS\\
11 & - & - & - & - & 23.10\,$\pm$\,0.22 & 6198.7\,$\pm$\,15.8 & 164.53\,$\pm$\,0.28 & 100.0 & BS\\
12 & - & - & - & - & 17.49\,$\pm$\,0.15 & 7307.9\,$\pm$\,10.2 & 155.30\,$\pm$\,0.08 & 72.3 & BS\\
13 & - & - & - & - & 20.92\,$\pm$\,0.23 & 7436.4\,$\pm$\,11.6 & 154.97\,$\pm$\,0.10 & 100.0 & BS\\
14 & - & - & - & - & 19.59\,$\pm$\,0.15 & 7778.4\,$\pm$\,11.0 & 157.70\,$\pm$\,0.08 & 98.8 & BS\\
15 & - & - & - & - & 23.60\,$\pm$\,0.50 & 2984.7\,$\pm$\,23.2 & 105.22\,$\pm$\,0.30 & 89.0 & BS\\
\end{tabular}
\justify
\textbf{Notes.} Column 1 is the label in the images shown in Fig. \ref{irdis2015-2016}, Cols. 2 and 3 are respectively the H2 and H3 relative magnitudes of point sources, Col. 4 is the angular separation to the star of the point sources for the first epoch, Col. 5 is the position angle of the point sources (0\deg\, is north and it is counted positive from north to east) for the first epoch, Col. 6 is the BB\_H relative magnitude of point sources, Col. 7 is the angular separation to the star of the point sources for the second epoch, Col. 8 is the position angle of the point sources for the second epoch, Col. 9 is the probability of each candidate to be a contaminant background star, based on the Besançon Galaxy Model \citep{2003A&A...409..523R}, Col. 10 is the current status of candidates (BS: background star).
\normalsize
\label{tabcc}
\end{table*}

\end{document}